\documentclass[9pt,conference]{IEEEtran}
\IEEEoverridecommandlockouts

\usepackage{cite}
\usepackage{amsmath,amssymb,amsfonts}
\usepackage{algorithmic}
\usepackage{graphicx}
\usepackage{subcaption}
\usepackage{textcomp}
\usepackage{xcolor}
\usepackage{svg}
\usepackage{tipa}
\usepackage{colortbl}
\usepackage{multirow,arydshln}
\usepackage{tablefootnote}
\usepackage{hyperref}
\usepackage{balance}

\usepackage{xcolor,soul}

\def\BibTeX{{\rm B\kern-.05em{\sc i\kern-.025em b}\kern-.08em
    T\kern-.1667em\lower.7ex\hbox{E}\kern-.125emX}}
\begin{document}

\title{Cross-Lingual Speech Emotion Recognition: \\ Humans vs. Self-Supervised Models}

\makeatletter
\newcommand{\linebreakand}{%
  \end{@IEEEauthorhalign}
  \hfill\mbox{}\par
  \mbox{}\hfill\begin{@IEEEauthorhalign}
}
\makeatother

\author{\IEEEauthorblockN{Zhichen Han}
\IEEEauthorblockA{\textit{University of Edinburgh, UK}}
\and
\IEEEauthorblockN{Tianqi Geng}
\IEEEauthorblockA{\textit{Tianjin University, China}}
\and
\IEEEauthorblockN{Hui Feng}
\IEEEauthorblockA{\textit{Tianjin University, China}}
\linebreakand
\IEEEauthorblockN{Jiahong Yuan}
\IEEEauthorblockA{\textit{University of Science and Technology of China, China}}
\and
\IEEEauthorblockN{Korin Richmond}
\IEEEauthorblockA{\textit{University of Edinburgh, UK}}
\and
\IEEEauthorblockN{Yuanchao Li$^\dag$\thanks{$^\dag$Corresponding author. \textit{yuanchao.li@ed.ac.uk}}}
\IEEEauthorblockA{\textit{University of Edinburgh, UK}}
}


\maketitle

\begin{abstract}
Utilizing Self-Supervised Learning (SSL) models for Speech Emotion Recognition (SER) has proven effective, yet limited research has explored cross-lingual scenarios. This study presents a comparative analysis between human performance and SSL models, beginning with a layer-wise analysis and an exploration of parameter-efficient fine-tuning strategies in monolingual, cross-lingual, and transfer learning contexts. We further compare the SER ability of models and humans at both utterance- and segment-levels. Additionally, we investigate the impact of dialect on cross-lingual SER through human evaluation. Our findings reveal that models, with appropriate knowledge transfer, can adapt to the target language and achieve performance comparable to native speakers. We also demonstrate the significant effect of dialect on SER for individuals without prior linguistic and paralinguistic background. Moreover, both humans and models exhibit distinct behaviors across different emotions. These results offer new insights into the cross-lingual SER capabilities of SSL models, underscoring both their similarities to and differences from human emotion perception.
\end{abstract}


\section{Introduction}
\label{sec:intro}

The advancement of Self-Supervised Learning (SSL) has led to the development of powerful pre-trained models, such as Wav2vec 2.0 (W2V2) \cite{baevski_wav2vec_2020} and WavLM \cite{chen_wavlm_2022}, including their multilingual variants. These models have demonstrated remarkable success across a range of downstream speech tasks, including Speech Emotion Recognition (SER) \cite{li2022fusing}. To further enhance their adaptability across different languages and datasets for SER, Parameter-Efficient Fine-Tuning (PEFT) has been utilized to improve the efficacy of SSL models while minimizing fine-tuning requirements \cite{feng_2023_peft,lashkarashvili_parameter_2024}.

Nevertheless, cross-lingual SER remains a significant challenge due to language and cultural differences \cite{shoukat_breaking_2023}. Typically, both traditional and SSL models require sufficient training data in the target language to achieve satisfactory cross-lingual SER performance, which is often infeasible for languages lacking emotional speech datasets \cite{latif_cross_2018, ahn_cross-corpus_2021}. For humans, however, although cross-lingual barriers exist \cite{elfenbein_universality_2002}, emotions in speech are universally distinguishable as humans are less affected by cross-lingual differences \cite{jeon_preliminary_2013}.

While some research has explored the use of SSL models for cross-lingual and multilingual SER \cite{singh2023decoding}, there has been little investigation into how these models compare to human performance. To this end, we raise four key questions:

\textit{\textbf{1)} Can SSL-based models achieve competitive SER performance to that of humans?}

\textit{\textbf{2)} How to better fine-tune SSL models for SER in cross-lingual scenarios?}

\textit{\textbf{3)} Does dialect have an impact on human perception in cross-lingual SER?}

\textit{\textbf{4)} Can SSL-based models identify emotionally salient segments similar to human behaviors?}

To answer the above questions, we conduct a comparative study between humans and SSL models, specifically:

\begin{itemize}
    \item We perform a layer-wise analysis and investigate various PEFT strategies for SSL models in monolingual, cross-lingual, and transfer learning settings, comparing SER performance with human performance across emotions.
    \item We evaluate SER performance on Tianjin speech (a Chinese dialect), exploring the impact of dialect on human listeners with and without linguistic and paralinguistic background knowledge.
    \item We assess both human and SSL model performance on the Speech Emotion Diarization (SED) task (i.e., segment-level SER), aiming to compare their ability to detect prominent emotion segments.
\end{itemize}




\section{Related Work}
\label{sec:relatedwork}

On the model side, previous studies have typically fine-tuned SER models using target language data, but have observed a significant drop in performance when shifting from monolingual to cross-lingual conditions \cite{latif_cross_2018,neumann_cross-lingual_2018}. Additionally, adversarial neural networks in unsupervised settings have been explored for cross-lingual adaptation \cite{latif_unsupervised_2019,cai_unsupervised_2021}. More recently, \cite{upadhyay2024layer} introduced a layer-anchoring mechanism to facilitate emotion transfer, accounting for the task-specific nature and hierarchical structure of speech models. On the human side, \cite{jeon_preliminary_2013} found that SVM models outperformed humans in monolingual settings, whereas humans were less affected by cross-lingual challenges. Further research by \cite{werner_speech_2019} concluded that human cross-lingual capabilities in SER are generally robust.

Despite this progress, comparative studies between humans and models remain lacking, leading to an insufficient understanding of human-model comparison. To our knowledge, we are the first to conduct a comparative study between humans and SSL models, exploring not only utterance-level SER but also fine-grained emotion perception (i.e., SED), the impact of dialect, and fine-tuning strategies.

\section{Materials and Methodology}
\label{sec:dataset}
\subsection{Datasets and Models}
As various tasks are investigated in this work, we use multiple datasets and models. For the datasets, four public emotion corpora and a non-public dialect corpus are used:

\begin{itemize}
    \item \textit{ESD}: a Mandarin Chinese (CN) emotion corpus \cite{zhou_emotional_2022}, containing utterances spoken by ten native CN speakers (five male, five female) across five emotion categories.
    \item \textit{PAVOQUE}: a German (DE) emotion corpus \cite{steiner_pavoque_2013}, featuring a professional male actor with five emotion categories, where neutral comprises over 50\% of the dataset.
    \item \textit{IEMOCAP}: an English (EN) emotion corpus \cite{busso_iemocap_2008}, where five male and female speakers were paired to record scripted and improvised emotional utterances, divided into nine emotion categories. 
    \item \textit{ZED}: an English emotion corpus specifically designed for the SED task \cite{wang2023speech}, with speech data annotated by humans at both utterance and sub-utterance (segment) levels.
    \item \textit{TJD}: a non-public Tianjin (TJ) Chinese dialect corpus collected in our previous work \cite{gengform}. It was recorded and annotated at Tianjin University by two native Tianjin dialect speakers. It includes three functional categories (\textit{question}, \textit{negation}, \textit{expectation}), approximated to emotions due to high acoustic similarity. According to annotators, \textit{negation} resembles \textit{anger}, and \textit{expectation} resembles \textit{happiness}. Tianjin dialect is known for its complex tone sandhi patterns while featuring a similar but slightly different tone system to Mandarin \cite{li2019tianjin}. Native speakers of the Tianjin dialect convey emotions more directly with noticeable sonorous vowels and faster speech \cite{qi_2020}.
\end{itemize}

For the models, we use three \textit{W2V2} base models pre-trained on Mandarin CN\footnote{\href{https://huggingface.co/TencentGameMate/chinese-wav2vec2-base}{https://huggingface.co/TencentGameMate/chinese-wav2vec2-base}}, DE\footnote{\href{https://huggingface.co/facebook/wav2vec2-base-de-voxpopuli-v2}{https://huggingface.co/facebook/wav2vec2-base-de-voxpopuli-v2}}, and EN\footnote{\href{https://huggingface.co/facebook/wav2vec2-base-960h}{https://huggingface.co/facebook/wav2vec2-base-960h}}, along with a \textit{WavLM} large model trained on EN emotional speech \footnote{\href{https://huggingface.co/speechbrain/emotion-diarization-wavlm-large}{https://huggingface.co/speechbrain/emotion-diarization-wavlm-large}}. The following tasks are conducted using different models and datasets for specific purposes.

\subsection{Layer-wise Analysis of SSL Models}

In this task, we use the \textbf{datasets}: \textit{ESD}, \textit{PAVOQUE}, \textit{IEMOCAP}; the \textbf{models}: \textit{W2V2-CN, -DE, -EN}; and the \textbf{emotions}: \textit{angry}, \textit{happy}, \textit{neutral}, and \textit{sad}.

SSL models encode speech information across different layers; specifically, in SER tasks, speech representations from the middle layers often yield higher performance \cite{li2023exploration}. Therefore, we perform a layer-wise analysis to identify the optimal layer for monolingual and cross-lingual SER. SSL models are used as feature extractors with all parameters frozen, and Unweighted Accuracy (UA) is used as the evaluation metric. The analysis is conducted in the following settings:

\begin{itemize}
    \item \textit{Monolingual} (Mono): The model is fine-tuned with both training and test data from speech in the same language as its pre-training language. For example, \textit{W2V2-CN} is fine-tuned using CN data (\textit{ESD}) as both training and test data.
    \item \textit{Cross-lingual} (Cross): The model is fine-tuned using its pre-training language as training data and a different language as test data. For example, \textit{W2V2-CN} is fine-tuned using CN data (\textit{ESD}) and tested on DE data (\textit{PAVOQUE}) or EN data (\textit{IEMOCAP}).
    \item \textit{Transfer learning} (Trans): The model is fine-tuned and tested on a language different from its pre-training language. For example, \textit{W2V2-CN} is fine-tuned and tested on either DE data (\textit{PAVOQUE}) or EN data (\textit{IEMOCAP}).
\end{itemize}



\subsection{PEFT of SSL Models for Cross-Lingual SER}

In this task, we use the \textbf{datasets}: \textit{ESD}, \textit{PAVOQUE}, \textit{IEMOCAP}; the \textbf{models}: \textit{W2V2-CN, -DE}; and the \textbf{emotions}: \textit{angry}, \textit{happy}, \textit{neutral}, and \textit{sad}.

After the layer-wise analysis, the best-performing layers are further fine-tuned using various PEFT strategies to enhance performance. We apply the Low-Rank Adapter (LoRA) \cite{hu_lora_2021}, Bottleneck Adapter (BA) \cite{houlsby_parameter-efficient_2019}, and Weighted Gating (WG) \cite{lashkarashvili_parameter_2024}. Additionally, a two-stage fine-tuning \cite{lashkarashvili_parameter_2024} is performed: the model is first fine-tuned on the source language, then on the target language once the first fine-tuning converges.

\subsection{Comparison of SSL Models with Human Evaluation}

In this task, we use all the datasets and models. For \textbf{SER}, we use the emotions: \textit{angry}, \textit{happy}, \textit{neutral}, and \textit{sad}; while for \textbf{SED}, we exclude \textit{neutral} as it does not contain emotion variation to perceive and segment.

Six native DE speakers (one male, five female) and six native CN speakers (two male, four female), with no prior knowledge of each other’s language, are recruited for the human evaluation from the University of Edinburgh and Tianjin University. All participants have studied English for many years with sufficient skills (e.g., IELTS score $\ge$ 6.5). The webMUSHRA interface \cite{schoeffler_webmushra_2018} is used to create the experimental tests.

For \textbf{SER}, participants listen to speech samples and identify the conveyed emotion. We use UA as the evaluation metric, consistent with the model performance evaluation. Additionally, to investigate fine-grained speech emotion expression, we perform \textbf{SED}, where participants first listen to speech samples and label the emotion, as in the SER task. Subsequently, they clip the speech and select the segment that most prominently expresses the emotion. Following \cite{wang2023speech}, we use the Emotion Diarization Error Rate (EDER) as the metric, which calculates the error rate of diarization results, including missed emotions (ME), false alarms (FA), overlaps (OL), and confusion (CF):

\begin{equation}
    EDER=\frac{ME+FA+OL+CF}{Uttrance\ Duration}
\end{equation}

For comparison with the SSL models, we compare participants' performance on their native language with the monolingual setting, their performance on the non-native languages with the cross-lingual or transfer learning settings. Finally,
we explore whether dialect has an impact on human perception of cross-lingual SER.



\section{Experiments}
\subsection{Experimental Settings}

For \textbf{SER}, to reduce the effect of varying training data sizes, we use the same amount of data for CN, DE, and EN. To ensure a balanced emotion distribution, we use an equal number of samples for each emotion. Specifically, for \textit{ESD}, \textit{PAVOQUE}, and \textit{IEMOCAP}, we apply 5-fold cross-validation for model training: 400 utterances per emotion category, totaling 1,600 utterances per dataset, are used for training. Similarly, 200 utterances are randomly selected for validation and test sets, respectively. Given the difficulty of performing human evaluation on all the data, for comparison with human evaluation, we select 12 sentences per emotion category, totaling 144 utterances for all languages (12 sentences $\times$ 4 emotions $\times$ 3 datasets). The model settings are as follows:

\subsubsection{Layer-wise analysis} 
We use a classification head projecting from dimension 768 to 4 for SER, with a learning rate of 1e-4, epsilon of 1e-8, and weight decay of 1e-5, trained for 100 epochs with a batch size of 32. Cross-entropy is used as the loss criterion. Training stops if the validation loss does not decrease for 10 consecutive epochs.

\subsubsection{PEFT strategies}
We use the same classification head configuration as in the layer-wise analysis for PEFT. For the LoRA module, the attention head is set to 8, alpha for scaling is 16, with a dropout rate of 0.1. For the BA module, the reduction factor is 16. Models are trained for 100 epochs with a batch size of 16. The loss and stopping criteria from the layer-wise analysis remain the same.

For \textbf{SED}, given the considerable effort required for segmenting speech, only 8 utterances per emotion are randomly selected from \textit{ZED}, totaling 24 utterances, for comparison with human evaluation and model results\footnote{Code available: \href{https://github.com/zhan7721/Crosslingual_SER}{https://github.com/zhan7721/Crosslingual\_SER}}.

\subsection{Results and Discussions}


\begin{figure}
    \centering
    \includegraphics[width=\columnwidth]{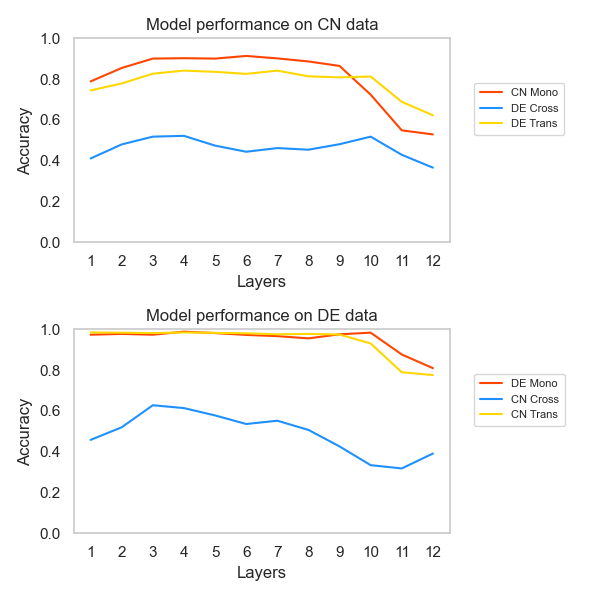}
    \caption{Layer-wise analysis of CN and DE models under monolingual, cross-lingual and transfer learning settings.}
    \label{fig:layerwise_analysis}
\end{figure}


The results of the layer-wise analysis are presented in Figure~\ref{fig:layerwise_analysis}. In the monolingual setting, both the CN and DE models demonstrate strong performance on their respective source languages, as expected, given that the models are pre-trained on these languages. However, in the cross-lingual setting, both models show a significant drop in accuracy. While this is reasonable due to language differences, the extent of the drop is beyond our expectations, considering the shared characteristics of emotional acoustics \cite{scherer2003vocal,banse1996acoustic}. One possible explanation is that SSL models not only encode low-level acoustic features but also transform them into high-level, linguistically related information, such as word identity and meaning \cite{pasad_layer-wise_2022}. This process creates a linguistic gap across languages, exacerbating the accuracy decline. Nonetheless, under the transfer learning setting, the models can achieve performance levels comparable to the monolingual setting, demonstrating the ability of SSL models to adapt to different languages for SER with appropriate techniques of knowledge transfer. The variations in the contours are related to the training objectives of SSL models, particularly the contrastive masked segment prediction (since these patterns align with previous research on layer-wise analysis of SSL models \cite{pasad_layer-wise_2022,li2023exploration,saliba2024layer}, we omit further detailed explanation).

The results of \textbf{PEFT} under monolingual, cross-lingual, and transfer learning settings, are shown in Table~\ref{tab:model_acc}. \textbf{Human performance} on SER is shown in Table~\ref{tab:human_acc}, and \textbf{SED comparison} is presented in Table~\ref{tab:SED_result}. From these results, we make the following observations:

\begin{table}[ht]
    \centering
    \caption{Model performance under monolingual, cross-lingual, and transfer learning with various PEFT strategies.}
    \scalebox{0.9}{
    \begin{tabular}{llllcccc}
    \hline
    \multirow{2}{*}{\textbf{Model}} & \multirow{2}{*}{\textbf{Setting}} & \multirow{2}{*}{\textbf{Source}} & \multirow{2}{*}{\textbf{Target}} & \multicolumn{3}{c}{\textbf{PEFT strategy}} & \multicolumn{1}{c}{\multirow{2}{*}{\textbf{UA\%}}} \\
     &  &  &  & \textbf{LoRA} & \textbf{BA+WG} & \textbf{2-stg} & \multicolumn{1}{c}{} \\
    \hline
        \multirow{13}{*}{\begin{tabular}[c]{@{}l@{}}\textit{W2V2}\\ \textit{-CN}\end{tabular}} 
            & \multirow{3}{*}{Mono} & \multirow{3}{*}{CN} & \multirow{3}{*}{CN} 
            &  &  &  & 91.4 \\
            & & & & \checkmark &  &  & 87.0 \\
            & & & & \checkmark & \checkmark &  & \textbf{93.9} \\
        \cdashline{2-8}
            & \multirow{3}{*}{Cross} & \multirow{3}{*}{CN} & \multirow{3}{*}{DE} 
            &  &  &  & 62.8 \\
            & & & & \checkmark &  &  & 65.3 \\
            & & & & \checkmark & \checkmark &  & \textbf{70.7} \\
        \cdashline{2-8}   
            & \multirow{4}{*}{Trans} & \multirow{4}{*}{DE} & \multirow{4}{*}{DE} 
            &  &  &  & 98.5 \\
            & & & & \checkmark &  &  & 98.8 \\
            & & & & \checkmark & \checkmark &  & 98.8 \\
            & & & & \checkmark & \checkmark & \checkmark & \textbf{98.9} \\
        \cdashline{2-8}
            & \multirow{3}{*}{Trans} & \multirow{3}{*}{EN} & \multirow{3}{*}{EN} 
            &  &  &  & 65.5 \\
            & & & & \checkmark &  &  & 66.3 \\
            & & & & \checkmark & \checkmark &  & \textbf{67.7} \\ 
    \hline
        \multirow{13}{*}{\begin{tabular}[c]{@{}l@{}}\textit{W2V2}\\ \textit{-DE}\end{tabular}} 
            & \multirow{3}{*}{Mono} & \multirow{3}{*}{DE} & \multirow{3}{*}{DE} 
            &  &  &  & \textbf{98.9} \\
            & & & & \checkmark &  &  & 97.8 \\
            & & & & \checkmark & \checkmark &  & 97.9 \\
        \cdashline{2-8}
            & \multirow{3}{*}{Cross} & \multirow{3}{*}{DE} & \multirow{3}{*}{CN} 
            &  &  &  & 52.2 \\
            & & & & \checkmark &  &  & \textbf{58.5} \\
            & & & & \checkmark & \checkmark &  & 56.0 \\
        \cdashline{2-8}
            & \multirow{4}{*}{Trans} & \multirow{4}{*}{CN} & \multirow{4}{*}{CN} 
            &  &  &  & 84.2 \\
            & & & & \checkmark &  &  & 83.6 \\
            & & & & \checkmark & \checkmark &  & \textbf{87.5} \\
            & & & & \checkmark & \checkmark & \checkmark & 85.8 \\
        \cdashline{2-8}
            & \multirow{3}{*}{Trans} & \multirow{3}{*}{EN} & \multirow{3}{*}{EN} 
            &  &  &  & 62.4 \\
            & & & & \checkmark &  &  & 65.0 \\
            & & & & \checkmark & \checkmark &  & \textbf{66.0 }\\ 
    \hline
    \end{tabular}
    }
    \label{tab:model_acc}
\end{table}

\begin{table}[ht]
    \centering
    \large
    \caption{Human performance (UA\%) on all languages. The higher the value, the better the SER performance.}
    \scalebox{0.9}{
    \begin{tabular}{ccccc}
    \hline
         & \textbf{CN} & \textbf{DE} & \textbf{EN} & \textbf{TJ} \\
    \hline
        \textbf{CN participants} & 79.5 & 73.3 & 63.5 & 67.5 \\
        \textbf{DE participants} & 82.6 & 91.7 & 73.6 & 29.2 \\
    \hline
    \end{tabular}}
    \label{tab:human_acc}
\end{table}

\textbf{\textit{1) SER: monolingual model vs.\ native speakers}}

In terms of overall accuracy, as shown in Table~\ref{tab:model_acc} and Table~\ref{tab:human_acc}, both models outperform their respective human native speakers. For predictions across all emotion categories, Table~\ref{tab:monoling_conf_matrix} presents the confusion matrices of the CN and DE monolingual models alongside those of CN and DE natives for their respective languages. Compared to the DE monolingual model, DE natives are more likely to report false alarms for \textit{sad} in \textit{neutral} DE speech. CN natives, compared to the CN monolingual model, demonstrate lower precision in \textit{happy} and \textit{neutral}. These results indicate that SSL models exhibit excellent monolingual performance on the SER task when provided with sufficient training data. 

\begin{table*}[ht]
\centering
\caption{Confusion matrices of CN and DE speakers and models under monolingual (first row), cross-lingual (second row) and transfer learning (third row) settings. No humans under the transfer learning setting.}
\begin{minipage}{0.45\columnwidth}
\centering
\scalebox{0.85}{
\begin{tabular}{ccccc}
\hline
      & \textbf{A} & \textbf{H} & \textbf{N} & \textbf{S} \\ \hline
    \textbf{A} & 0.71 & 0.24 & 0.06 & 0.00 \\
    \textbf{H} & 0.00 & 0.82 & 0.18 & 0.00 \\
    \textbf{N} & 0.00 & 0.10 & 0.86 & 0.04 \\
    \textbf{S} & 0.00 & 0.01 & 0.18 & 0.81 \\
\hline
\end{tabular}}
\caption*{\footnotesize{(a) Human: CN (mono)}}
\end{minipage}%
\hfill
\begin{minipage}{0.45\columnwidth}
\centering
\scalebox{0.85}{
\begin{tabular}{ccccc}
\hline
      & \textbf{A} & \textbf{H} & \textbf{N} & \textbf{S} \\ \hline
    \textbf{A} & 0.95 & 0.05 & 0.00 & 0.00 \\
    \textbf{H} & 0.02 & 0.97 & 0.02 & 0.00 \\
    \textbf{N} & 0.00 & 0.00 & 1.00 & 0.00 \\
    \textbf{S} & 0.00 & 0.00 & 0.03 & 0.97 \\
\hline
\end{tabular}}
\caption*{\footnotesize{(b) Model: CN mono (on CN)}}
\end{minipage}%
\hfill
\begin{minipage}{0.45\columnwidth}
\centering
\scalebox{0.85}{
\begin{tabular}{ccccc}
\hline
      & \textbf{A} & \textbf{H} & \textbf{N} & \textbf{S} \\ \hline
    \textbf{A} & 0.97 & 0.03 & 0.00 & 0.00 \\
    \textbf{H} & 0.00 & 0.94 & 0.05 & 0.00 \\
    \textbf{N} & 0.00 & 0.00 & 0.83 & 0.17 \\
    \textbf{S} & 0.00 & 0.00 & 0.04 & 0.96 \\
\hline
\end{tabular}}
\caption*{\footnotesize{(c) Human: DE (mono)}}
\end{minipage}%
\hfill
\begin{minipage}{0.45\columnwidth}
\centering
\scalebox{0.85}{
\begin{tabular}{ccccc}
\hline
      & \textbf{A} & \textbf{H} & \textbf{N} & \textbf{S} \\ \hline
    \textbf{A} & 0.98 & 0.02 & 0.00 & 0.00 \\
    \textbf{H} & 0.02 & 0.98 & 0.00 & 0.00 \\
    \textbf{N} & 0.00 & 0.00 & 0.93 & 0.07 \\
    \textbf{S} & 0.00 & 0.00 & 0.00 & 1.00 \\
\hline
\end{tabular}}
\caption*{\footnotesize{(d) Model: DE mono (on DE)}}
\end{minipage}

\begin{minipage}{0.45\columnwidth}
\centering
\scalebox{0.85}{
\begin{tabular}{ccccc}
\hline
      & \textbf{A} & \textbf{H} & \textbf{N} & \textbf{S} \\ \hline
    \textbf{A} & 0.81 & 0.13 & 0.06 & 0.01 \\ 
    \textbf{H} & 0.21 & 0.38 & 0.42 & 0.00 \\ 
    \textbf{N} & 0.01 & 0.01 & 0.93 & 0.04 \\ 
    \textbf{S} & 0.01 & 0.00 & 0.15 & 0.83 \\
\hline
\end{tabular}}
\caption*{\footnotesize{(e) Human: CN (cross on DE)}}
\end{minipage}%
\hfill
\begin{minipage}{0.45\columnwidth}
\centering
\scalebox{0.85}{
\begin{tabular}{ccccc}
\hline
      & \textbf{A} & \textbf{H} & \textbf{N} & \textbf{S} \\ \hline
    \textbf{A} & 0.95 & 0.02 & 0.03 & 0.00 \\ 
    \textbf{H} & 0.08 & 0.18 & 0.51 & 0.22 \\ 
    \textbf{N} & 0.00 & 0.00 & 0.95 & 0.05 \\ 
    \textbf{S} & 0.00 & 0.00 & 0.00 & 1.00 \\
\hline
\end{tabular}}
\caption*{\footnotesize{(f) Model: CN cross (on DE)}}
\end{minipage}%
\hfill
\begin{minipage}{0.45\columnwidth}
\centering
\scalebox{0.85}{
\begin{tabular}{ccccc}
\hline
      & \textbf{A} & \textbf{H} & \textbf{N} & \textbf{S} \\ \hline
    \textbf{A} & 0.88 & 0.07 & 0.06 & 0.00 \\
    \textbf{H} & 0.08 & 0.75 & 0.15 & 0.01 \\ 
    \textbf{N} & 0.03 & 0.03 & 0.76 & 0.18 \\ 
    \textbf{S} & 0.00 & 0.03 & 0.06 & 0.92 \\
\hline
\end{tabular}}
\caption*{\footnotesize{(g) Human: DE (cross on CN)}}
\end{minipage}%
\hfill
\begin{minipage}{0.45\columnwidth}
\centering
\scalebox{0.85}{
\begin{tabular}{ccccc}
\hline
      & \textbf{A} & \textbf{H} & \textbf{N} & \textbf{S} \\ \hline
    \textbf{A} & 0.52 & 0.48 & 0.00 & 0.00 \\ 
    \textbf{H} & 0.17 & 0.65 & 0.17 & 0.02 \\ 
    \textbf{N} & 0.00 & 0.38 & 0.38 & 0.23 \\ 
    \textbf{S} & 0.05 & 0.28 & 0.30 & 0.37 \\
\hline
\end{tabular}}
\caption*{\footnotesize{(h) Model: DE cross (on CN)}}
\end{minipage}


\begin{minipage}{0.45\columnwidth}
\centering
\scalebox{0.85}{
\begin{tabular}{ccccc}
\hline
      & \textbf{A} & \textbf{H} & \textbf{N} & \textbf{S} \\ \hline
    \textbf{A} & 0.54 & 0.22 & 0.13 & 0.11 \\ 
    \textbf{H} & 0.14 & 0.61 & 0.19 & 0.06 \\ 
    \textbf{N} & 0.04 & 0.28 & 0.58 & 0.10 \\ 
    \textbf{S} & 0.03 & 0.01 & 0.15 & 0.81 \\
\hline
\end{tabular}}
\caption*{\footnotesize{(k) Human: CN (L2 on EN)}}
\end{minipage}%
\hfill
\begin{minipage}{0.45\columnwidth}
\centering
\scalebox{0.85}{
\begin{tabular}{ccccc}
\hline
      & \textbf{A} & \textbf{H} & \textbf{N} & \textbf{S} \\ \hline
    \textbf{A} & 0.70 & 0.13 & 0.10 & 0.07 \\ 
    \textbf{H} & 0.08 & 0.57 & 0.35 & 0.00 \\ 
    \textbf{N} & 0.07 & 0.18 & 0.75 & 0.00 \\ 
    \textbf{S} & 0.00 & 0.05 & 0.02 & 0.93 \\
\hline
\end{tabular}}
\caption*{\footnotesize{(l) Model: CN trans (on EN)}}
\end{minipage}%
\hfill
\begin{minipage}{0.45\columnwidth}
\centering
\scalebox{0.85}{
\begin{tabular}{ccccc}
\hline
      & \textbf{A} & \textbf{H} & \textbf{N} & \textbf{S} \\ \hline
    \textbf{A} & 0.78 & 0.03 & 0.04 & 0.15 \\ 
    \textbf{H} & 0.06 & 0.69 & 0.22 & 0.03 \\ 
    \textbf{N} & 0.15 & 0.13 & 0.60 & 0.13 \\ 
    \textbf{S} & 0.01 & 0.01 & 0.10 & 0.88 \\
\hline
\end{tabular}}
\caption*{\footnotesize{(m) Human: DE (L2 on EN)}}
\end{minipage}%
\hfill
\begin{minipage}{0.45\columnwidth}
\centering
\scalebox{0.85}{
\begin{tabular}{ccccc}
\hline
      & \textbf{A} & \textbf{H} & \textbf{N} & \textbf{S} \\ \hline
    \textbf{A} & 0.70 & 0.12 & 0.17 & 0.02 \\ 
    \textbf{H} & 0.00 & 0.65 & 0.33 & 0.02 \\ 
    \textbf{N} & 0.08 & 0.30 & 0.58 & 0.03 \\ 
    \textbf{S} & 0.00 & 0.00 & 0.25 & 0.75 \\
\hline
\end{tabular}}
\caption*{\footnotesize{(n) Model: DE trans (on EN)}}
\end{minipage}
\vspace{-10pt}
\label{tab:monoling_conf_matrix}
\end{table*}

\textbf{\textit{2) SER: cross-lingual models vs. humans}}

In terms of overall accuracy, as shown in Table~\ref{tab:model_acc} and Table~\ref{tab:human_acc}, both humans and models experience a performance decrease in the cross-lingual condition, with cross-lingual models being more significantly affected than humans. This aligns with findings from \cite{jeon_preliminary_2013}, which demonstrated that humans are capable of handling cross-lingual scenarios better. In terms of performance on every emotion category, as shown in Table~\ref{tab:monoling_conf_matrix}, DE cross-lingual model struggles to recognize \textit{neutral} and \textit{sad} in CN data, exhibiting low recall. Additionally, the DE model confuses \textit{angry} and \textit{happy} more frequently compared to humans in both languages. Conversely, the CN cross-lingual model closely aligns with CN natives when recognizing DE speech, with both often predicting \textit{happy} as \textit{neutral}.

Moreover, we conduct a two-sided Welch’s t-test on humans’ precision, recall, and F1-scores. We notice significant difference in the recall of \textit{happy} on DE data between CN and DE speakers ($t(10)=-7.511, p<0.001$), as well as in the precision of \textit{neutral} ($t(10)=-5.614, p<0.001$). CN speakers also exhibit lower recall for \textit{happy} in DE data than in CN data ($t(10)=-5.137, p<0.001$), suggesting a linguistic and paralinguistic knowledge gap between two speaker groups. Particularly, significant differences are found in the recall of \textit{sad} across CN, DE, and EN data ($t(10)=-2.708, p=0.022$) and in the precision of \textit{neutral} ($t(10)=-7.511, p<0.001$). The precision of \textit{neutral} is largely impacted by CN speakers' difficulty in perceiving \textit{happy} in DE data, indicating that linguistic and paralinguistic differences affect the perception of \textit{sad} across languages.

\textbf{\textit{3) SER: transfer learning models vs. L2 learners}}

As the transfer learning setting resembles the human learning process of a second language (i.e., fine-tuning $\approx$ language study), we compare the models with human speakers using EN data. As shown in Table~\ref{tab:model_acc}, SSL models with transfer learning achieve monolingual-level performance and surpass human accuracy on CN and DE data. However, for EN data, DE speakers exhibit higher accuracy than CN speakers and all models tested on EN data. Additionally, two-stage fine-tuning does not result in a significant performance boost, which was observed in the cross-corpus scenario under the same language \cite{lashkarashvili_parameter_2024}. These findings suggest that while transfer learning helps SSL models in adapting to new languages, performance varies depending on the specific target language dataset. In terms of performance on every emotion category, shown in Table~\ref{tab:monoling_conf_matrix}, CN speakers only outperform the model in recognizing \textit{happy}, whereas the CN transfer learning model outperforms humans in the other three emotion categories. For DE speakers, humans perform better at predicting \textit{happy} and \textit{neutral} compared to the DE transfer learning model. In addition, an effective PEFT strategy used in monolingual scenarios is not necessarily useful in cross-lingual or multilingual scenarios.

Moreover, Table~\ref{tab:human_acc} reveals that recognizing emotion in EN is more challenging than in CN and DE, despite CN and DE speakers being L2 learners. This difficulty is likely attributed to the selection of only improvised utterances from IEMOCAP, which are more natural and real-life emotions, thus making SER more challenging.

\textbf{\textit{4) SER: linguistic and paralinguistic impact of dialect}}

In addition to the finding in Observation 2 that linguistic and paralinguistic differences impact emotion perception across languages, the results on the TJ data in Table~\ref{tab:human_acc} further indicate the existence of such differences, particularly due to dialect. The SER results demonstrate the generalizability of human emotion perception across languages. However, in the \textit{TJD} dataset, performance varies significantly between the two speaker groups. While DE speakers excel with CN speech data, the unique prosody of the TJ dialect leads to a notable performance decline among DE speakers. This discrepancy is plausible given that TJ prosody and tones differ significantly from CN (and likely many other major languages), making emotion recognition challenging for DE speakers. Even with some background knowledge, CN speakers also struggle to recognize emotions in TJ data as effectively as in CN data, confirming the linguistic and paralinguistic impact of dialect.

\textbf{\textit{5) SED: models vs. humans in prominent emotion perception}}

The results in Table~\ref{tab:SED_result} indicate that both human groups outperform the model, with the DE speakers achieving the lowest EDER. The model performs best on \textit{happy} and worst on \textit{sad}. Between the human groups, CN speakers are slightly better at perceiving \textit{angry} segments, while DE speakers are better at identifying \textit{sad} segments. This pattern is consistent with SER results in Table~\ref{tab:monoling_conf_matrix}, where CN speakers show a higher threshold for predicting \textit{sad}, leading to higher recall but lower precision. Conversely, DE speakers demonstrate higher precision but lower recall. The difference in sensitivity to \textit{sad} among CN speakers results in more false negatives for \textit{sad} in the SED task.

\begin{table}[ht]
    \centering
    \large
    \caption{EDER (\%) comparison of WavLM and humans on ZED data. The lower the score, the better the performance.}
    \scalebox{0.9}{
    \begin{tabular}{ccccc}
    \hline
         & WavLM & CN participants & DE participants \\
    \hline
        Angry & 36.6 & \textbf{25.8} & 27.5 \\
        Happy & \textbf{27.5} & 31.8 & 28.7 \\
        Sad & 50.3 & 38.6 & \textbf{28.3} \\
    \hdashline
        Average & 38.2 & 32.1 & \textbf{28.2} \\
    \hline
    \end{tabular}}
    \label{tab:SED_result}
\end{table}

\section{Conclusion}

In this study, we conduct a comparative analysis of cross-lingual SER between humans and SSL models, including both modeling and human experiments, and compare their performance in monolingual, cross-lingual, and transfer learning settings. We perform a layer-wise analysis and apply PEFT to the best-performing layers using multiple strategies to enhance model performance. Additionally, we implement SED for fine-grained detection of salient emotion segments to evaluate the ability of SSL models to capture segment-level emotion. The results show that humans excel in cross-lingual SER and SED, while models can adapt to the target language through transfer learning to achieve native speaker-level performance. We also reveal the linguistic and paralinguistic impact of dialect in the cross-lingual setting through human evaluations. Our study provides novel insights into human emotion perception and the application of SSL models for cross-lingual SER.

\balance
\bibliographystyle{IEEEbib}
\bibliography{refs}

\begin{thebibliography}{10}

\bibitem{baevski_wav2vec_2020}
Alexei Baevski, Yuhao Zhou, Abdelrahman Mohamed, and Michael Auli,
\newblock ``wav2vec 2.0: A framework for self-supervised learning of speech representations,''
\newblock {\em Advances in neural information processing systems}, vol. 33, pp. 12449--12460, 2020.

\bibitem{chen_wavlm_2022}
Sanyuan Chen, Chengyi Wang, Zhengyang Chen, Yu~Wu, Shujie Liu, Zhuo Chen, Jinyu Li, Naoyuki Kanda, Takuya Yoshioka, Xiong Xiao, et~al.,
\newblock ``Wavlm: Large-scale self-supervised pre-training for full stack speech processing,''
\newblock {\em IEEE Journal of Selected Topics in Signal Processing}, vol. 16, no. 6, pp. 1505--1518, 2022.

\bibitem{li2022fusing}
Yuanchao Li, Peter Bell, and Catherine Lai,
\newblock ``Fusing {ASR} outputs in joint training for speech emotion recognition,''
\newblock in {\em ICASSP 2022-2022 IEEE International Conference on Acoustics, Speech and Signal Processing (ICASSP)}. IEEE, 2022, pp. 7362--7366.

\bibitem{feng_2023_peft}
Tiantian Feng and Shrikanth Narayanan,
\newblock ``{PEFT-SER}: On the use of parameter efficient transfer learning approaches for speech emotion recognition using pre-trained speech models,''
\newblock in {\em 2023 11th International Conference on Affective Computing and Intelligent Interaction (ACII)}. IEEE, 2023, pp. 1--8.

\bibitem{lashkarashvili_parameter_2024}
Nineli Lashkarashvili, Wen Wu, Guangzhi Sun, and Philip~C Woodland,
\newblock ``Parameter efficient finetuning for speech emotion recognition and domain adaptation,''
\newblock in {\em ICASSP 2024-2024 IEEE International Conference on Acoustics, Speech and Signal Processing (ICASSP)}. IEEE, 2024, pp. 10986--10990.

\bibitem{shoukat_breaking_2023}
Moazzam Shoukat, Muhammad Usama, Hafiz~Shehbaz Ali, and Siddique Latif,
\newblock ``Breaking barriers: Can multilingual foundation models bridge the gap in cross-language speech emotion recognition?,''
\newblock in {\em 2023 Tenth International Conference on Social Networks Analysis, Management and Security (SNAMS)}. IEEE, 2023, pp. 1--9.

\bibitem{latif_cross_2018}
Siddique Latif, Adnan Qayyum, Muhammad Usman, and Junaid Qadir,
\newblock ``Cross lingual speech emotion recognition: Urdu vs. western languages,''
\newblock in {\em 2018 International conference on frontiers of information technology (FIT)}. IEEE, 2018, pp. 88--93.

\bibitem{ahn_cross-corpus_2021}
Youngdo Ahn, Sung~Joo Lee, and Jong~Won Shin,
\newblock ``Cross-corpus speech emotion recognition based on few-shot learning and domain adaptation,''
\newblock {\em IEEE Signal Processing Letters}, vol. 28, pp. 1190--1194, 2021.

\bibitem{elfenbein_universality_2002}
Hillary~Anger Elfenbein and Nalini Ambady,
\newblock ``On the universality and cultural specificity of emotion recognition: a meta-analysis.,''
\newblock {\em Psychological bulletin}, vol. 128, no. 2, pp. 203, 2002.

\bibitem{jeon_preliminary_2013}
Je~Hun Jeon, Duc Le, Rui Xia, and Yang Liu,
\newblock ``A preliminary study of cross-lingual emotion recognition from speech: automatic classification versus human perception.,''
\newblock in {\em Interspeech}, 2013, pp. 2837--2840.

\bibitem{singh2023decoding}
Anant Singh and Akshat Gupta,
\newblock ``Decoding emotions: A comprehensive multilingual study of speech models for speech emotion recognition,''
\newblock {\em arXiv preprint arXiv:2308.08713}, 2023.

\bibitem{neumann_cross-lingual_2018}
Michael Neumann et~al.,
\newblock ``Cross-lingual and multilingual speech emotion recognition on {English} and {French},''
\newblock in {\em ICASSP 2018-2018 IEEE International Conference on Acoustics, Speech and Signal Processing (ICASSP)}. IEEE, 2018, pp. 5769--5773.

\bibitem{latif_unsupervised_2019}
Siddique Latif, Junaid Qadir, and Muhammad Bilal,
\newblock ``Unsupervised adversarial domain adaptation for cross-lingual speech emotion recognition,''
\newblock in {\em 2019 8th international conference on affective computing and intelligent interaction (ACII)}. IEEE, 2019, pp. 732--737.

\bibitem{cai_unsupervised_2021}
Xiong Cai, Zhiyong Wu, Kuo Zhong, Bin Su, Dongyang Dai, and Helen Meng,
\newblock ``Unsupervised cross-lingual speech emotion recognition using domain adversarial neural network,''
\newblock in {\em 2021 12th International Symposium on Chinese Spoken Language Processing (ISCSLP)}. IEEE, 2021, pp. 1--5.

\bibitem{upadhyay2024layer}
Shreya~G Upadhyay, Carlos Busso, and Chi-Chun Lee,
\newblock ``A layer-anchoring strategy for enhancing cross-lingual speech emotion recognition,''
\newblock {\em ICASSP 2024 - 2024 IEEE International Conference on Acoustics, Speech and Signal Processing (ICASSP)}, 2024.

\bibitem{werner_speech_2019}
Stefan Werner and Georgii~K Petrenko,
\newblock ``Speech emotion recognition: humans vs machines,''
\newblock {\em Discourse}, vol. 5, no. 5, pp. 136--152, 2019.

\bibitem{zhou_emotional_2022}
Kun Zhou, Berrak Sisman, Rui Liu, and Haizhou Li,
\newblock ``Emotional voice conversion: Theory, databases and esd,''
\newblock {\em Speech Communication}, vol. 137, pp. 1--18, 2022.

\bibitem{steiner_pavoque_2013}
Ingmar Steiner, Marc Schr{\"o}der, and Annette Klepp,
\newblock ``The {PAVOQUE} corpus as a resource for analysis and synthesis of expressive speech,''
\newblock {\em Proc. Phonetik \& Phonologie}, vol. 9, 2013.

\bibitem{busso_iemocap_2008}
Carlos Busso, Murtaza Bulut, Chi-Chun Lee, Abe Kazemzadeh, Emily Mower, Samuel Kim, Jeannette~N Chang, Sungbok Lee, and Shrikanth~S Narayanan,
\newblock ``{IEMOCAP}: Interactive emotional dyadic motion capture database,''
\newblock {\em Language resources and evaluation}, vol. 42, pp. 335--359, 2008.

\bibitem{wang2023speech}
Yingzhi Wang, Mirco Ravanelli, and Alya Yacoubi,
\newblock ``Speech emotion diarization: Which emotion appears when?,''
\newblock in {\em 2023 IEEE Automatic Speech Recognition and Understanding Workshop (ASRU)}. IEEE, 2023, pp. 1--7.

\bibitem{gengform}
Tianqi Geng and Hui Feng,
\newblock ``Form and function in prosodic representation: In the case of ‘ma’in tianjin mandarin,''
\newblock in {\em Interspeech}, 2024.

\bibitem{li2019tianjin}
Qian Li, Yiya Chen, and Ziyu Xiong,
\newblock ``Tianjin mandarin,''
\newblock {\em Journal of the International Phonetic Association}, vol. 49, no. 1, pp. 109--128, 2019.

\bibitem{qi_2020}
Shuling Qi,
\newblock {\em A Study of Tianjin Dialect's Grammar},
\newblock Shanghai Jiaotong University Press, 2020.

\bibitem{li2023exploration}
Yuanchao Li, Yumnah Mohamied, Peter Bell, and Catherine Lai,
\newblock ``Exploration of a self-supervised speech model: A study on emotional corpora,''
\newblock in {\em 2022 IEEE Spoken Language Technology Workshop (SLT)}. IEEE, 2023, pp. 868--875.

\bibitem{hu_lora_2021}
Edward~J Hu, Phillip Wallis, Zeyuan Allen-Zhu, Yuanzhi Li, Shean Wang, Lu~Wang, Weizhu Chen, et~al.,
\newblock ``{LoRA}: Low-rank adaptation of large language models,''
\newblock in {\em International Conference on Learning Representations}, 2021.

\bibitem{houlsby_parameter-efficient_2019}
Neil Houlsby, Andrei Giurgiu, Stanislaw Jastrzebski, Bruna Morrone, Quentin De~Laroussilhe, Andrea Gesmundo, Mona Attariyan, and Sylvain Gelly,
\newblock ``Parameter-efficient transfer learning for nlp,''
\newblock in {\em International conference on machine learning}. PMLR, 2019, pp. 2790--2799.

\bibitem{schoeffler_webmushra_2018}
Michael Schoeffler, Sarah Bartoschek, Fabian-Robert St{\"o}ter, Marlene Roess, Susanne Westphal, Bernd Edler, and J{\"u}rgen Herre,
\newblock ``{webMUSHRA}—a comprehensive framework for web-based listening tests,''
\newblock 2018.

\bibitem{scherer2003vocal}
Klaus~R Scherer,
\newblock ``Vocal communication of emotion: A review of research paradigms,''
\newblock {\em Speech communication}, vol. 40, no. 1-2, pp. 227--256, 2003.

\bibitem{banse1996acoustic}
Rainer Banse and Klaus~R Scherer,
\newblock ``Acoustic profiles in vocal emotion expression.,''
\newblock {\em Journal of personality and social psychology}, vol. 70, no. 3, pp. 614, 1996.

\bibitem{pasad_layer-wise_2022}
Ankita Pasad, Ju-Chieh Chou, and Karen Livescu,
\newblock ``Layer-wise analysis of a self-supervised speech representation model,''
\newblock in {\em 2021 IEEE Automatic Speech Recognition and Understanding Workshop (ASRU)}. IEEE, 2021, pp. 914--921.

\bibitem{saliba2024layer}
Alexandra Saliba, Yuanchao Li, Ramon Sanabria, and Catherine Lai,
\newblock ``Layer-wise analysis of self-supervised acoustic word embeddings: A study on speech emotion recognition,''
\newblock in {\em 2024 IEEE International Conference on Acoustics, Speech, and Signal Processing Workshops (ICASSPW)}. IEEE, 2024.

\end{thebibliography}

\end{document}